\documentclass[conference]{IEEEtran}
\IEEEoverridecommandlockouts
\usepackage{cite}
\usepackage{amsmath,amssymb,amsfonts}
\usepackage{algorithmic}
\usepackage{graphicx}
\usepackage{textcomp}
\usepackage{xcolor}

\usepackage{hyperref}
\usepackage{multirow}
\usepackage{stfloats}
\usepackage{booktabs}

\def\BibTeX{{\rm B\kern-.05em{\sc i\kern-.025em b}\kern-.08em
    T\kern-.1667em\lower.7ex\hbox{E}\kern-.125emX}}
\begin{document}

\title{M2R-Whisper: Multi-stage and Multi-scale Retrieval Augmentation for Enhancing Whisper\\
\thanks{$^{*}$Corresponding author. This work has been supported by the National Key R$\&$D Program of China (Grant No.2022ZD0116307) and NSF China (Grant No.62271270)}
}

\author{\IEEEauthorblockN{Jiaming Zhou$^1$, Shiwan Zhao$^1$, Jiabei He$^1$, Hui Wang$^1$,
Wenjia Zeng$^2$, \\ Yong Chen$^2$,
Haoqin Sun$^1$, Aobo Kong$^1$ and Yong Qin$^{1,*}$}
\IEEEauthorblockA{$^1$TMCC, College of Computer Science, Nankai University, Tianjin, China \\
$^2$Lingxi (Beijing) Technology Co., Ltd.\\
Email: zhoujiaming@mail.nankai.edu.cn}
}

\maketitle

\begin{abstract}

State-of-the-art models like OpenAI's Whisper exhibit strong performance in multilingual automatic speech recognition (ASR), but they still face challenges in accurately recognizing diverse subdialects. In this paper, we propose M2R-Whisper, a novel multi-stage and multi-scale retrieval augmentation approach designed to enhance ASR performance in low-resource settings. Building on the principles of in-context learning (ICL) and retrieval-augmented techniques, our method employs sentence-level ICL in the pre-processing stage to harness contextual information, while integrating token-level $k$-Nearest Neighbors ($k$NN) retrieval as a post-processing step to further refine the final output distribution. By synergistically combining sentence-level and token-level retrieval strategies, M2R-Whisper effectively mitigates various types of recognition errors. Experiments conducted on Mandarin and subdialect datasets, including AISHELL-1 and KeSpeech, demonstrate substantial improvements in ASR accuracy, all achieved without any parameter updates.

\end{abstract}


\section{Introduction}

The rapid advancements in deep learning have led to remarkable progress in automatic speech recognition (ASR) systems \cite{prabhavalkar2023end, Conformer, espnet, zhang2022wenet, cif-t, whisper, usm, mms}, resulting in models like OpenAI’s Whisper \cite{whisper}, Google’s USM \cite{usm}, and META’s MMS \cite{mms} that support multilingual speech recognition and achieve near state-of-the-art performance. However, these models still encounter difficulties in low-resource settings \cite{wang2024enhancing,slt-wsy, madi,childmandarin, stutter}, such as recognizing diverse subdialects \cite{basak2023challenges,zhu2023boosting,kothawade2023ditto,aksenova2022accented}. These limitations highlight the need for more flexible and adaptive methods to enhance ASR performance under such conditions.

Against this backdrop, in-context learning (ICL) \cite{rubin2022learning,min2022rethinking} has emerged as a promising approach in natural language processing (NLP), allowing models to leverage relevant examples embedded within the input, enabling adaptation without finetuning.
The success of ICL hinges on the quality of selected demonstrations \cite{agrawal2023context}, which poses unique challenges in speech processing due to the complexity of audio data. Existing approaches \cite{gao22e_interspeech, hsu2023exploration} often rely on random selection, leading to suboptimal results. For instance, WAVPROMPT \cite{gao22e_interspeech} combines Wav2vec2 with an autoregressive language model for few-shot ICL in speech tasks, while Hsu et al. \cite{hsu2023exploration} propose a text-free warmup strategy to enable ICL in speech models. However, both methods are limited to classification tasks, leaving ICL applications in ASR relatively underexplored.

Recent studies have begun exploring the application of ICL in ASR, particularly in low-resource settings. For instance, SLAM \cite{chen2024salm} leverages large language models (LLMs) to enhance ASR performance by improving keyword recognition through ICL, while Audio Flamingo \cite{audioflamingo} enables fast adaptation to unseen tasks using sentence-level embeddings. However, Audio Flamingo requires an additional audio encoder, such as LAION-CLAP \cite{LAION-CLAP}, for retrieval. While these approaches utilize the ICL capabilities of LLMs, Wang et al. \cite{prompt-whisper} explore Whisper’s ICL potential for test-time adaptation in Chinese dialects, leveraging small labeled speech samples to reduce the Character Error Rate (CER). Nonetheless, their approach is limited to isolated word-level ICL, which constrains its applicability to broader ASR contexts.

\begin{figure}[!t]
  \centering
  \includegraphics[width=\linewidth]{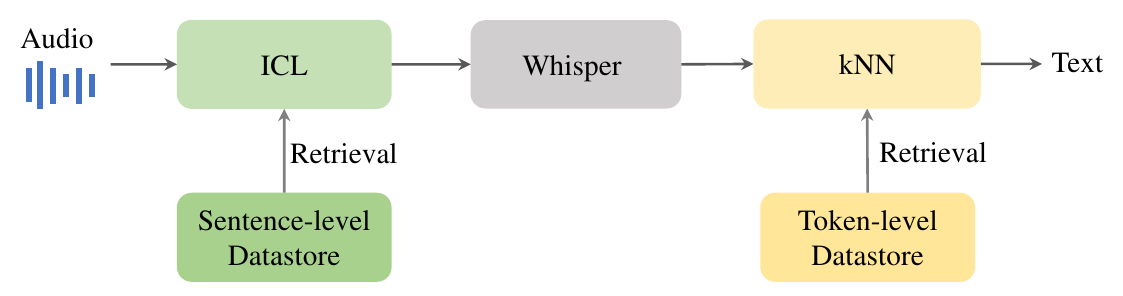}
  \caption{
  An overview of M2R-Whisper: a multi-stage and multi-scale retrieval augmentation framework. This framework enhances the Whisper ASR model by incorporating sentence-level retrieval in the pre-processing stage and token-level retrieval in the post-processing stage, with the goal of improving recognition accuracy, particularly in low-resource subdialect settings.}
  \label{intro}
\end{figure}

\begin{figure*}[!t]
  \centering
  \includegraphics[width=\linewidth]{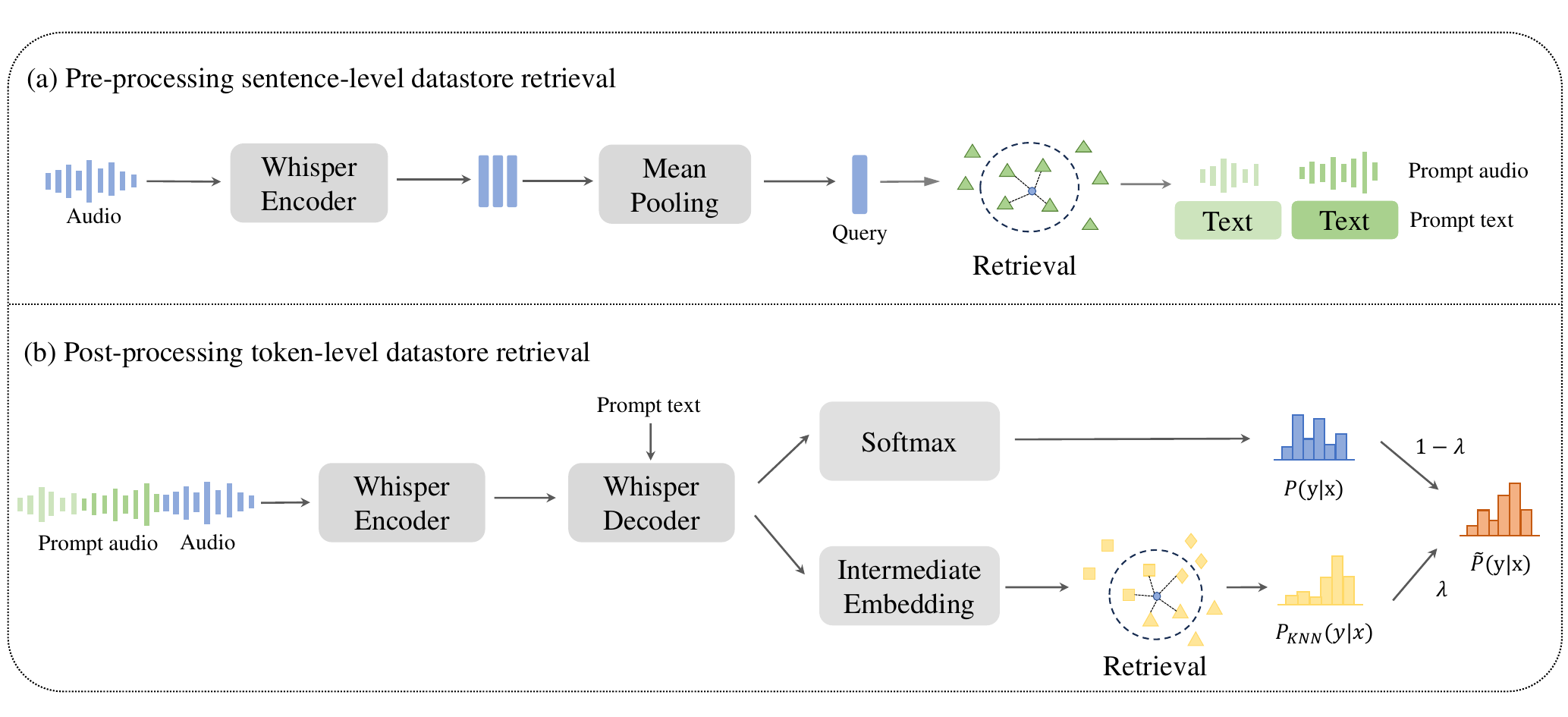}
  \caption{
  Illustration of our M2R-Whisper framework, which integrates multi-stage and multi-scale retrieval augmentation. The method consists of pre-processing with sentence-level ICL and post-processing with token-level $k$NN. Prior to testing, we construct separate sentence-level and token-level datastores from the training set. For each test audio, we retrieve the top $k$ most similar audio-text pairs as prompts, concatenating the prompt audio with the test audio. The corresponding prompt text is passed as a special token prefix to enhance ICL. During decoding, token-level retrieval augmentation is applied to generate the $k$NN distribution $P_{kNN}$, which is then interpolated with Whisper's output distribution $P$ to produce the final prediction.
}
  \label{overview}
\end{figure*}

Beyond ICL, retrieval-augmented methods \cite{knn-lm,knn-mt,FAISS,xu2023nearest}, have been widely adopted in NLP to enhance model performance without parameter updating. Techniques like $k$NN-LM \cite{knn-lm} for language modeling and $k$NN-MT \cite{knn-mt} for machine translation have shown significant promise. In ASR, $k$NN-CTC \cite{knn-ctc} uses Connectionist Temporal Classification (CTC) pseudo labels to create speech-text key-value pairs and retrieves these labels during decoding to refine the output distribution, significantly improving performance in Chinese dialect ASR. Building on $k$NN-CTC, Zhou et al. \cite{knn-cs} further extend this approach with a gated mechanism for zero-shot code-switching ASR. However, these methods are constrained by the quality of the pseudo labels, and $k$NN-CTC primarily addresses substitution errors, showing instability in correcting other types of errors.

While both ICL and $k$NN have demonstrated significant success in enhancing ASR performance, they differ substantially in their operational stages, retrieval scales, methods of leveraging external datastores, and the types of ASR errors they primarily address. By integrating these complementary approaches, we can achieve synergistic improvements in ASR. Based on this insight, we propose a multi-stage and multi-scale retrieval augmentation method for test-time adaptation, specifically tailored for Whisper, as illustrated in Figure \ref{intro}.

Our approach extends word-level ICL \cite{prompt-whisper} to sentence-level ICL, enhancing Whisper’s in-context learning capabilities. While ICL serves as a pre-processing retrieval mechanism that leverages the model's intrinsic ability, its impact remains limited. To address this, we integrate post-processing token-level $k$NN retrieval to further refine the final output distribution, forming a multi-stage and multi-scale retrieval augmentation ASR system.
On the other hand, while token-level $k$NN retrieval primarily focuses on reducing substitution errors, sentence-level retrieval effectively handles a broader range of errors. Combining these methods yields complementary effects, improving ASR’s ability to address diverse errors across different stages and scales.

This approach facilitates rapid domain adaptation without requiring parameter updates. Extensive experiments on AISHELL-1 \cite{aishell} and KeSpeech \cite{kespeech} validate the effectiveness of our method, demonstrating that the integration of sentence-level and token-level datastores enhances ASR performance.

Our main contributions are as following:

1. We extend word-level ICL to sentence-level for Whisper by establishing a sentence-level datastore.

2. We implement a token-level datastore to realize the $k$NN retrieval augmentation method for Whisper.

3. We propose a multi-stage and multi-scale retrieval augmentation method, M2R-Whisper, which achieves complementary gains by combining both sentence-level and token-level approaches, without updating any parameters.

4. Comprehensive experiments demonstrate the effectiveness of our method on both Mandarin and subdialect datasets.

\section{Our method}

Our method consists of two key components: pre-processing sentence-level ICL and post-processing token-level $k$NN. 
Combining both components results in complementary effects, enhancing different performance aspects of the ASR system. Our method is illustrated in Figure \ref{overview}. 

\subsection{Pre-processing sentence-level ICL}
Inspired by the word-level ICL approach for Whisper \cite{prompt-whisper}, we extend this to the sentence level for use as a pre-processing retrieval mechanism. This section outlines how we construct a sentence-level datastore using the training set $S$ and retrieve the top $k$ most similar audio-text pairs to prompt Whisper, thereby enhancing ICL performance. 

\textbf{Datastore construction}:
Given a Whisper model and an input audio $X$ from $S$, we extract the encoder output embeddings and apply mean pooling across the frames to obtain $f(X)$ as the key.
The corresponding ground-truth label $Y$ serves as the value, forming a sentence-level datastore $D_{s}$:

\begin{equation}
    D_{s}=\{(f(X),Y)| X \in S\},
\end{equation}

\textbf{Prompt retrieval}:
During testing, the query $f(X)$ is derived from the test audio, which is used to retrieve the top $k$ nearest audio-text pairs from the datastore. These retrieved pairs are employed to prompt the Whisper model to harness contextual information, with the goal of improving ASR performance.

\textbf{Performing ICL for Whisper}:
Whisper provides a special tokens $prefix$ \footnote{More details could be found at \url{https://github.com/openai/whisper/discussions/117\#discussioncomment-3727051}} to accept partial text of the input audio, helping manage long audio inputs (over 30 seconds). We concatenate the retrieved prompt audios with the current testing audio and input the ground-truth text label as the special token $prefix$ to boost Whisper's ICL capabilities.

\subsection{Post-processing token-level $k$NN}
Inspired by $k$NN-CTC \cite{knn-ctc}, we introduce a token-level datastore for Whisper, using ground-truth tokens instead of frame-level CTC pseudo labels.

\textbf{Datastore construction}: For each input audio $X$ in the training set $S$, we construct a token-level datastore $D_t$ by:
\begin{equation}
    D_{t}=(K,V)=\{(g(x_i),y_i)| X \in S\},
\end{equation}
where $g(x_i)$ is the $i$th intermediate embedding extracted from Whisper, specifically the input to the final decoding layer's feed-forward network (FFN) after layer normalization. $y_i$ is the corresponding ground truth token.

\textbf{Candidate retrieval}:
During decoding, we extract the intermediate embedding $g(x)$ as a query to retrieve $k$ nearest neighbours $\mathcal{N}$ at each step.
The $k$NN distribution is then computed by aggregating the probabilities of each vocabulary unit based on the retrieved neighbors $\mathcal{N}$ as follows:
\begin{equation}
P_{kNN}(y|x) \propto \sum_{(K_i, V_i)\in\mathcal{N},V_i=y} {exp(-d(K_i,g(x)/\tau))},
\end{equation}
where $K_i$ and $V_i$ are the $i$-th key and corresponding value respectively, $\tau$ denotes the temperature, $d(\cdot,\cdot)$ is the $L^2$ distance. 
For simplicity, speech and text prompts are omitted from this equation.

The final prediction is obtained by interpolating Whisper's output distribution with the $k$NN distribution:
\begin{equation}
\widetilde{P}(y|x)=\lambda P_{kNN}(y|x)+(1-\lambda)P(y|x),
\end{equation}
where $\lambda$ is a hyperparamter to balance the Whisper output distribution and $k$NN distribution.

\section{Experimental setup}
\subsection{Dataset}
The details of each dataset we employ are shown in Table \ref{dataset}.
We conducted our experiments using two datasets:

\textbf{AISHELL-1}: AISHELL-1 \cite{aishell} is a widely-used open-source ASR corpus consisting of 178 hours of Mandarin speech data from 400 speakers.

\textbf{KeSpeech}: KeSpeech \cite{kespeech} is an open-source speech dataset containing 1,542 hours of Chinese speech, including eight subdialects. We utilized four Chinese subdialect ASR subsets from KeSpeech: Jiang-Huai, Ji-Lu, Zhongyuan, and Southwestern.

\subsection{Baselines}
The following method are considered for comparison:
\begin{itemize}
\item {\bf Whisper}: We use the Large-V2 version of Whisper \cite{whisper} as our baseline. Other methods incorporating Whisper are also based on the same version.
\item {\bf $k$NN-CTC} :
This method \cite{knn-ctc} uses a frame-level datastore to enhance the performance of the Conformer-CTC model during greedy search. 
The model is trained on 10,000 hours of labeled Mandarin speech data.
We utilize the full version of $k$NN-CTC for comparison.

\item {\bf $k$NN-Whisper}: This approach employs a token-level datastore to perform retrieval augmentation for Whisper.

\item {\bf Prompt-Whisper}:
This method extends word-level in-context learning (ICL) \cite{prompt-whisper} to the sentence level for Whisper by creating a sentence-level datastore to retrieve similar audio samples for ICL.

\item {\bf M2R-Whisper}: This is our proposed multi-stage and multi-scale retrieval augmentation method for Whisper, combining both sentence-level and token-level retrieval strategies.
\end{itemize}

\begin{table}[t]
\centering
\caption{Dataset details}
\label{dataset}
\begin{tabular}{ccccc}
\toprule
 Dataset        & Subdialect        & \# Hours & \# Speakers              & \# Utterances \\
\midrule
AISHELL-1       &Mandarin         & 178    & 400                     & 141,600      \\
\midrule
\multirow{4}{*}{KeSpeech}        &Jiang-Huai              & 46     & 2,913                   & 30,008       \\
                &Ji-Lu                   & 59     & 2,832                   & 36,921       \\
                &ZhongYuan              & 84     & 3,038                   & 52,012       \\
                &Southwestern             & 75     & 2,646                   & 48,465       \\
\bottomrule
\end{tabular}
\end{table}

\subsection{Implementation details}

We employ Whisper Large-V2 as our baseline, which contains 1,550 million parameters and is trained on 630,000 hours of web-scraped speech data, demonstrating excellent performance on multilingual ASR tasks.
We set retrieval neighbours $k$ to 16 for both sentence-level and token-level candidate retrieval processes. 
In practice, the number of prompts $n$ is capped at 10, discarding any additional retrieved samples due to Whisper's 30-second input limit.
For token-level $k$NN retrieval augmentation, the optimal value for the hyperparameter $\lambda$ is typically between 0.2 and 0.4, determined by tuning on the development set.
For sentence-level candidate retrieval, we concatenate audio samples up to a maximum duration of 30 seconds to achieve the best performance
All the reported results are based on greedy search decoding to ensure a fair comparison.

For the Whisper Large-V2 baseline, many Simplified Chinese characters are transcribed as Traditional, and digits as Arabic numerals instead of their pronunciations, leading to a higher CER. To ensure fair comparison, we post-process the Whisper results by converting Traditional to Simplified Chinese and replacing Arabic numerals with their Chinese pronunciations. Notably, these inconsistencies minimally affect our method, demonstrating its stability and effectiveness.

\begin{table*}[th]
\caption{CER (\%) and Relative CER Reduction (RR) (\%) of different methods on Mandarin and Mandarin subdialect datasets}
\label{main-res}
\centering
\begin{tabular}{c|cc|cc|cc|cc|cc|cc} 
\toprule
\multirow{2}{*}{Method} & \multicolumn{2}{c|}{AISHELL-1}  & \multicolumn{2}{c|}{Ji-Lu}       & \multicolumn{2}{c|}{Jiang-Huai}  & \multicolumn{2}{c|}{ZhongYuan}   & \multicolumn{2}{c|}{Southwestern} & \multicolumn{2}{c}{Avg.}        \\

                        & CER $\downarrow$         & RR $\uparrow$    & CER $\downarrow$         & RR $\uparrow$    & CER $\downarrow$         & RR $\uparrow$  & CER $\downarrow$         & RR $\uparrow$    & CER $\downarrow$         & RR $\uparrow$  & CER $\downarrow$         & RR $\uparrow$                  \\
\midrule
Whisper                 & 5.76          & -              & 31.31          & -              & 40.96          & -              & 32.57          & -              & 31.57           & -              & 28.43          & -              \\
$k$NN-CTC            & 4.78          & 17.01          & 28.61          & 8.62           & 39.95          & 2.47           & 30.04          & 7.77           & 28.24           & 10.55          & 26.32          & 7.42           \\
$k$NN-Whisper             & 4.41          & 25.52          & 26.75          & 14.56          & 37.62          & 9.15           & 25.18            & 22.69          & 28.59           & 9.44          & 24.51          & 13.80          \\
Prompt-Whisper         & 4.77          & 19.27          & 27.77          & 11.30          & 37.27          & 9.01           & 27.42        & 15.81          & 24.00           & 23.98           & 24.25          & 14.73          \\
M2R-Whisper             & \textbf{4.11} & \textbf{30.73} & \textbf{24.11} & \textbf{22.99} & \textbf{35.62} & \textbf{13.04} & \textbf{23.26} & \textbf{28.58} & \textbf{21.43}  & \textbf{32.12} & \textbf{21.71} & \textbf{23.66} \\
\bottomrule
\end{tabular}
\end{table*}

\begin{figure}[!t]
  \centering
  \includegraphics[width=0.9\linewidth]{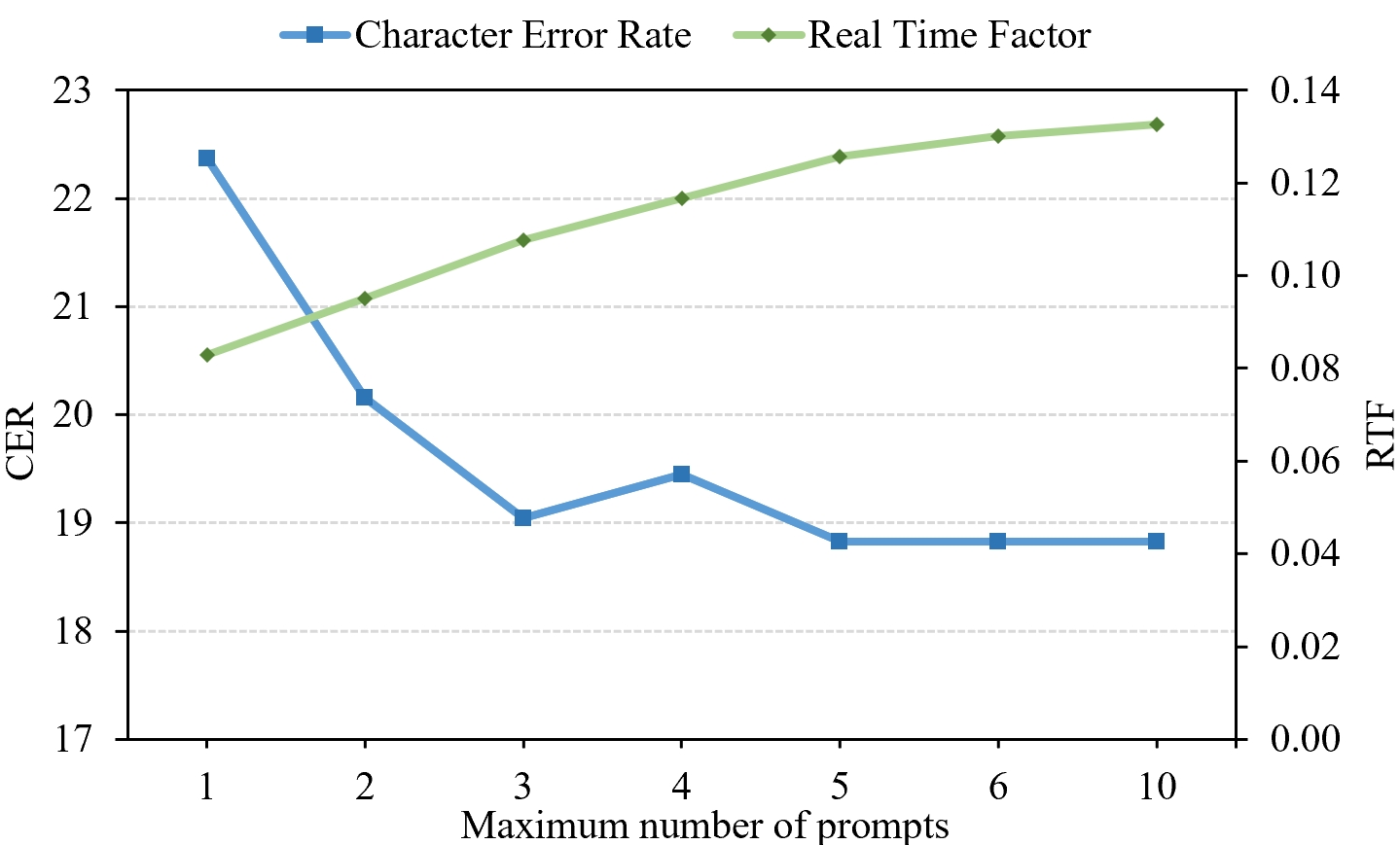}
  \caption{CER (\%) and RTF for different maximum numbers of prompts on the Southwestern development set}
  \label{CER_REF}
\end{figure}

\begin{table}[]
\caption{Quantities of Substitution (S), Deletion (D), and Insertion (I) errors on ZhongYuan and Southwestern subdialect datasets}
\label{error_analysis}
\centering
\begin{tabular}{c|c|ccc|c}
\toprule
Dataset                       & Method         & S      & D     & I   & CER $\downarrow$\\
\midrule

\multirow{4}{*}{Southwestern} & Whisper        & 10,464 & 809   & 842 & 31.57 \\
                              & KNN-Whisper    & 8,709  & 1,263 & 999 & 28.59 \\
                              & Prompt-Whisper & 8,165  & \textbf{472}   & 573 & 24.00 \\
                              & M2R-Whisper     & \textbf{7,065}  & 879   & \textbf{277} & \textbf{21.43} \\
\midrule
\multirow{4}{*}{Ji-Lu} & Whisper         & 10,907 & 957   & 1,000 & 31.31\\
                       & KNN-Whisper     & 8,714  & 1,722 & 395  & 26.75 \\
                       & Prompt-Whisper  & 10,052 & \textbf{508}   & 683   & 27.77\\
                       & M2R-Whisper      & \textbf{8,303}  & 1,104 & \textbf{357}  & \textbf{24.11}\\
\bottomrule
\end{tabular}
\end{table}

\section{Result}

Table \ref{main-res} presents the primary results, reporting the CER (\%) and relative CER reduction (\%) on the AISHELL-1 Mandarin dataset and four subdialect datasets. Whisper Large-V2 is used as the baseline, achieving a CER of 5.76\% on AISHELL-1 but showing significantly higher error rates on the subdialect datasets, highlighting the difficulties in low-resource ASR tasks involving subdialects.
Both $k$NN-Whisper and $k$NN-CTC outperform the Whisper baseline across all datasets, underscoring the efficacy of fine-grained retrieval augmentation techniques. Notably, Whisper's auto-regressive decoding allows for the construction of a more accurate datastore using ground truth labels rather than CTC pseudo labels. Consequently, $k$NN-Whisper surpasses $k$NN-CTC, suggesting that the quality of the datastore plays a pivotal role in performance improvement.
Prompt-Whisper achieves better results than $k$NN-Whisper on 3 out of the 5 datasets, with a relative CER reduction of 14.73\% compared to the baseline. This superior performance suggests that similar audio-text pairs provide enhanced contextual information, allowing Whisper to decode more accurately. Our proposed method, M2R-Whisper, achieves the best overall performance across all datasets without requiring any parameter updates. Notably, M2R-Whisper reaches a CER of 4.11\% on AISHELL-1, reflecting a relative reduction of 30.73\%, and an average relative reduction of 23.66\% across all datasets. These results demonstrate the effectiveness of our multi-stage and multi-scale retrieval augmentation approach.

In Figure \ref{CER_REF}, we further explore the impact of varying the maximum number of audio-text pairs used to prompt Whisper. As the number of prompts increases, CER consistently decreases. However, Real-Time Factor
\footnote{\url{https://openvoice-tech.net/index.php/Real-time-factor}} 
(RTF) results reveal that the ICL approach incurs a high RTF, indicating that concatenating multiple prompt audios can slow down the decoding process. Adjusting the maximum number of prompts helps balance decoding efficiency and performance.

To further explore the collaborative effects of both methods, we present a detailed breakdown of substitution (S), deletion (D), and insertion (I) errors in Table \ref{error_analysis}, facilitating a deeper analysis of each method's strengths and limitations.
We observe that $k$NN-Whisper primarily focuses on reducing substitution errors but tends to increase deletion errors, and its inconsistent handling of insertion errors in the Southwestern and Ji-Lu subsets indicates its instability in this regard.
In contrast, Prompt-Whisper effectively mitigates deletion and insertion errors, addressing the shortcomings introduced by $k$NN-Whisper.
Overall, M2R-Whisper delivers the fewest errors, especially in reducing substitution and insertion errors. This demonstrates that M2R-Whisper, by integrating the strengths of both $k$NN-Whisper and Prompt-Whisper, effectively enhances ASR performance, addressing diverse types of errors at multiple stages and scales.

\section{Conclusion}
In this paper, we introduced M2R-Whisper, a novel approach that integrates sentence-level ICL in the pre-processing stage and token-level $k$NN retrieval in the post-processing stage. This combination achieves complementary improvements without parameter updates. Extensive experiments on both Mandarin and Mandarin subdialect datasets confirm the effectiveness of our approach,in enhancing ASR performance.
\newpage

\bibliographystyle{IEEEbib}
\bibliography{ref}

\end{document}